\documentstyle[aps,amssymb,epsfig,twocolumn]{revtex}
\begin{document}
\textheight 230mm
\def\bff{ }
\def\bee{ }
\def\NNstructures{186 }
\def\NNprototypes{108 }
\def\NNprototypesPoint{108.}
\def\NNalloys{6 }
\def\NNalloysC{5 }
\def\NNalloysS{1 }
\def\NNcalc{14,080 }
\def\NNTOTcalc{32,402 }
\def\NNsecs{$\sim 9.05\cdot10^8$ }   
\def\NNyears{$\sim 28.7$ }           

\def\DEG{{$^\circ$C}}
\def\massalski{{Massalski \cite{BB}}}
\def\PawSection{as described in Section (\ref{section.firstprinciple})}
\def\us{us-lda}
\def\paw{paw-gga}
\def\tablefont{\footnotesize}
\def\tablelineone{{\it Ab initio}}\def\tablelinetwo{result} 
\def\twophase{$\leftrightarrow$}
\def\picdim{76mm}
\def\ftnsz{\footnotesize}

\wideabs
{
\title{\LARGE 
{\sf High-throughput ab initio analysis of the \\
Bi-In, Bi-Mg, Bi-Sb, In-Mg, In-Sb, and Mg-Sb systems }}
\author{
Stefano Curtarolo$^{1,2}$, Aleksey N. Kolmogorov$^{1}$, Franklin Hadley Cocks$^{1}$}
\address{
$^1$Department of Mechanical Engineering and Materials Science, Duke University, Durham, NC 27708 \\
$^2${corresponding author, e-mail: stefano@duke.edu}
}
\date{\today}
\maketitle
\begin{abstract}
{ Prediction and characterization of crystal structures of alloys are
  a key problem in materials research. Using high-throughput {\it ab
  initio} calculations we explore the low-temperature phase diagrams
  for the following systems: {Bi-In, Bi-Mg, Bi-Sb, In-Mg, In-Sb, and
  Mg-Sb}. For the experimentally observed phases in these systems we
  provide information about their stability at low temperatures.

  }
\end{abstract}
}


\section{Introduction}
\label{section.introduction}

Magnesium alloys present an excellent combination of light weight and
specific strength compared to steel and aluminum alloys and would
offer an opportunity for radical improvement in automotive vehicle
design and performance\cite{Hadley1,Hadley2}. However, for more than
fifty years magnesium alloy research has been confined almost
exclusively to casting alloys and there has been no equivalent effort
in developing wrought magnesium alloys of sufficiently high
strength\cite{Hadley3}. Both wrought aluminum alloys and wrought
magnesium alloys achieve their increased strength levels through the
process of precipitation-hardening, also known as age-hardening. In
the case of aluminum alloys, the metallurgy of precipitation-hardening
has been thoroughly explored and has resulted in the five and six
element aluminum alloys which form the basis for all aerospace
structures\cite{Hadley4}. The trial and error research has required an
enormous effort over many years to produce the aluminum alloys now in
use. This is so because the hardening process depends on the
development of stable and metastable phases in the base alloy matrix,
a problem particularly difficult for experimental investigation due to
long annealing times involved in the phase formation. If the
development of greatly increased strength in precipitation-hardenable
magnesium alloys followed a similar trial and error path it would
require a similar level of effort\cite{Hadley5}.

Fortunately, the combination of {\it ab initio} density functional
theory methods and data mining techniques provides an opportunity to
dramatically accelerate materials research by efficiently predicting
new phases and accurately describing their ground states
\cite{CurtaPRL2003,CurtaCALPHAD2005,MorganMRS2004,MorganJMST2004,WangCALPHAD2004}.
These theoretical methods are particularly suitable for investigation
of low-temperature compounds and can thus play an important role in
the development of precipitation-hardenable magnesium alloys.

In this paper we systematically explore the low-temperature phase
diagrams for the following six binary alloy systems: 
Bi-In, Bi-Mg, Bi-Sb, In-Mg, In-Sb, and Mg-Sb. 
Both Mg-Bi and Mg-Sb systems have the
potential to be new age-hardenable magnesium-based alloys. Indium is
of interest because it is one of the very few elements with high
solubility in magnesium and yet has low electrochemical activity;
whereby the relatively high electrochemical activity of magnesium
might reasonably be expected to be reduced, albeit at an increase in
overall alloy density. The In-Sb and In-Bi systems are included to
complete the possibilities for binary alloys among these elements.

The Binary Alloy Phase Diagrams\cite{BB} and the Pauling File\cite{PP}
give a broad review of experimental data on these systems. In most
cases the experimental results are complemented by thermodynamic
modeling \cite{BB,PP,CC}. However, to the best of our knowledge, {\it
ab initio} studies regarding phase stabilities in these systems are
scarce \cite{Wijs96,Guo93,Kel00}. To check the completeness of the
experimentally known phase diagrams in these systems and provide
information on the phase stabilities at low temperatures we have
chosen a large library of most common prototypes in binary alloys and
calculated ground state energies of these structures with {\it ab
initio} methods.

We describe the prototype library in the next section, give the
details of the {\it ab initio} methods used in this study in section
``High-throughput First Principles calculations'', and present the
results for each of the six systems in section ``Alloys''.

\section{Binary systems and structure prototypes}

We calculate six alloys, Bi-In, Bi-Mg, Bi-Sb, In-Mg, In-Sb, and Mg-Sb in
\NNstructures crystal structure configurations.  Many of these
configurations have the same prototype, for example, AB$_3$ and A$_3$B
so the number of distinct prototypes is \NNprototypesPoint\, The various
concentrations are listed in the following table.
 
\begin{center}   
  {\tablefont
    \begin{tabular}{||c|c|c||}  \hline  
      Compounds           & Conc.    & number of  \\ 
      composition         & of B     & prototypes  \\ \hline
         A \&          B  &  0\%     &  6 \\ \hline
    A$_5$B \&     AB$_5$  &  16.66\% &  3 \\ \hline
    A$_4$B \&     AB$_4$  &  20\%    &  2 \\ \hline
    A$_3$B \&     AB$_3$  &  25\%    & 27 \\ \hline 
A$_2$B$_5$ \& A$_5$B$_2$  &  28.57\% & 1 \\ \hline 
    A$_2$B \&     AB$_2$  &  33.33\% & 34 \\ \hline
A$_5$B$_3$ \& A$_3$B$_5$  &  37.5\%  &  3 \\ \hline
A$_3$B$_2$ \& A$_2$B$_3$  &  40\%    &  2 \\ \hline
A$_4$B$_3$ \& A$_3$B$_4$  &  42.85\% &  1 \\ \hline
        AB (\& BA$^*$)    &  50\%    & 29 (+3) \\ \hline
    \end{tabular}  
  }
\end{center}
    {TABLE 1.  Compositions, concentrations and number of prototypes
      in the library.  The library has \NNstructures structures, and
      \NNprototypes distinct prototypes ($^*$ at composition AB, 3
      prototypes have different point groups in atomic positions A and
      B, therefore they represent distinct structure types).  }
\vspace{2mm}

Of such prototypes, 67 are chosen from the most common intermetallic
binary structures in the Pauling File \cite{PP} and the {\small CRYSTMET} database \cite{CC},
plus the common low temperature compounds (5) reported for the Mg, In, Sb and Bi systems \cite{BB,PP}.
Such prototypes can be described by their Strukturbericht designation 
and/or natural prototypes \cite{BB,PP}:
A1, A2, A3, A4, A6, A7, A15, B$_{h}$, B1, B2, B3, B4, B8$_{1}$,
B8$_{2}$, B10, B11, B19, B27, B32, B33 (B$_{f}$), C$_{c}$, C2, C6,
C11$_{b}$, C14, C15, C15$_{b}$, C16, C18, C22, C32, C33, C37, C38,
C49, D0$_{a}$, D0$_{3}$, D0$_{9}$, D0$_{11}$, D0$_{19}$, D0$_{22}$,
D0$_{23}$, D0$_{24}$, D1$_{3}$, D1$_{a}$, D2$_{d}$, D5$_2$,
D7$_{3}$, D8$_{8}$, D8$_g$, L1$_{0}$, L1$_{1}$, L1$_{2}$,
L6$_{0}$, CaIn$_{2}$, Cr$_3$B$_5$, CuTe, CuZr$_{2}$, GdSi$_{2}$
(1.4), Mg$_2$In, MoPt$_{2}$, NbAs (NbP), NbPd$_{3}$, 
Ni$_2$In, Ni$_{2}$Si, $\Omega$ (with z=1/4), Pu$_{3}$Al (Co$_{3}$V),
Ti$_{3}$Cu$_{4}$, W$_{5}$Si$_{3}$, YCd$_{3}$, ZrSi$_{2}$, $\gamma$-Ir.
The rest of the structures (36) are fcc, bcc or hcp superstructures.
Twelve of these superstructures consist of stacking of pure A and B
planes along some common direction \cite{CurtaPRL2003,CurtaMIT2003}.

\section{High-throughput First Principles calculations}
\label{section.firstprinciple}

The high-throughput {\it ab initio} method used for this project is fully
described in references
\cite{CurtaPRL2003,CurtaCALPHAD2005}
A summary of the details of the calculations is given below.

{\bf Ultra Soft Pseudopotentials LDA calculations (US-LDA).}  The
energy calculations were performed using Density Functional Theory in
the Local Density Approximation (LDA), with the Ceperley-Alder form
for the correlation energy as parameterized by Perdew-Zunger
\cite{PerdewZunger} and with ultra soft Vanderbilt type
pseudopotentials \cite{Vanderbilt}, as implemented in {\small
VASP} \cite{VASP}.  Calculations are done at zero temperature and
pressure, with spin polarization, and without zero-point motion.  The
energy cutoff in an alloy was set to 1.5 times the larger of the
suggested energy cutoffs of the pseudopotentials of the elements of
the alloy (suggested energy cutoffs are derived by the method
described in \cite{VASP}).  Brillouin zone integrations were
performed using at least 3500/(number of atoms in unit cell) ${\bf
k}$-points distributed on a Monkhorst-Pack mesh
\cite{MONKHORST_PACK1,MONKHORST_PACK2}.  With these energy
cutoffs and ${\bf k}$-points meshes the absolute energy is converged
to better than 10 meV/atom.  Energy differences between structures are
expected to be converged to much smaller tolerances.  All structures
were fully relaxed. 

{\bf PAW-GGA calculations.}  When several structures are in close
competition for the ground state, we also performed calculations in
the Generalized Gradient Approximation (GGA), with Projector
Augmented-Wave (PAW) pseudopotentials, as implemented in {\small VASP}
\cite{VASP,PAW_BLOCK,VASP_PAW}.  In general, we expect the
PAW-GGA approach to be more accurate than the US-LDA.  For the GGA
correlation energy, we used the Perdew-Wang parameterization
(GGA-PW91) \cite{perdew_wang_PW91}. Compared to the US-LDA case,
we use an increased energy cutoff of 1.75 times the larger of the
suggested energy cutoffs for the elements in the system and a finer
${\bf k}$-point mesh with at least $\sim$4000/(number of atoms in unit
cell).

{\bf Symmetries of the pure elements.}  Our calculations reproduce
the correct experimental crystal structures of the pure elements at
room temperature.  Bi and Sb are most stable in the A7 structure
($\alpha$-As prototype), while Mg and In have hexagonal closed packed
(A3) and face-centered tetragonal (A6) structures, respectively.

{\bf Calculation of the formation energies and the convex hull.}  The
formation energy for each structure is determined with respect to the
most stable structure of the pure elements.  To determine the ground
states of a system one needs to find, as a function of composition,
the ordered compounds that have an energy lower than any other
structure or any linear combination of structures that gives the
proper composition. This set of ground state structures forms a {\it
convex hull}, as all other structures have an energy that falls above
the set of tie lines that connects the energy of the ground states.
In thermodynamical terms, the {\it convex hull} represents the Gibbs
free energy of the alloy at zero temperature.

\newpage
\section{Alloys}
\label{section.compoundforming}


\begin{center}{\bf Bi-In (Bismuth - Indium)}\end{center}  
The phase diagram of the Bi-In system is known from experimental
investigations and thermodynamic modeling
\cite{BB,PP,Gie67,Boo73,Deg82,Eva83,Che88,BiInSb,Can94}. Three phases
are expected to be stable at low temperatures: BiIn-B10,
BiIn$_2$-InNi$_2$, and Bi$_3$In$_5$-Cr$_3$B$_5$. However, the evidence
for the low-temperature stability of BiIn$_2$ and Bi$_3$In$_5$ is not
conclusive\cite{Boo73}.  With our US-LDA calculations, we have found
that only three phases have negative formation energy.  These phases
BiIn-B10, BiIn$_2$-InNi$_2$, and a monoclinic structure with
space group C2/m \#12 at 50\% concentration have formation energies
equal to -17 meV/atom, -4.4 meV/atom, and -7.9 meV/atom, respectively.
Therefore, phase diagram calculated in US-LDA has only one stable
compound, BiIn-B10; the other two experimentally observed stable phases
BiIn$_2$ and Bi$_3$In$_5$ are above the convex hull by 6.3 and 26
meV/atom, respectively.

{\tablefont
  \begin{center}
    \begin{tabular}{||c||} \hline
      {\bf Bi-In system} \\ \hline
      Low Temperature Phases comparison chart \\ \hline
      \begin{tabular}{c|c|c}
        Composition  & Experimental  & \tablelineone                       \\
        \% Bi        & (\massalski)  & \tablelinetwo                       \\ \hline
        33.3         &    BiIn$_2$-InNi$_2$     &   two phase region (\us) \\
        \            &                          &   BiIn$_2$-InNi$_2$ is   \\
        \            &                          &   $\sim$6.3 meV/at. higher than \\
        \            &                          &   the tie-line In \twophase BiIn \\
        \            &                          &   {\it BiIn$_2$ stable (\paw)} \\ \hline
        37.5         &    Bi$_3$In$_5$-Cr$_3$B$_5$  & Bi$_3$In$_5$ unstable\\
        \            &                          &   \us\,\, and {\it \paw}       \\  \hline
        50           &   BiIn-B10               &   BiIn-B10               \\
        \            &                          &   E$_f$=-17 meV/at. (\us) \\ 
        \            &                          &   {\it E$_f$=-6 meV/at. (\paw)} \\
      \end{tabular} \\ \hline
    \end{tabular}
  \end{center}
}

\vspace{-7mm}
\begin{center}
  \vspace{-3mm}
  \begin{figure}
    \epsfig{file=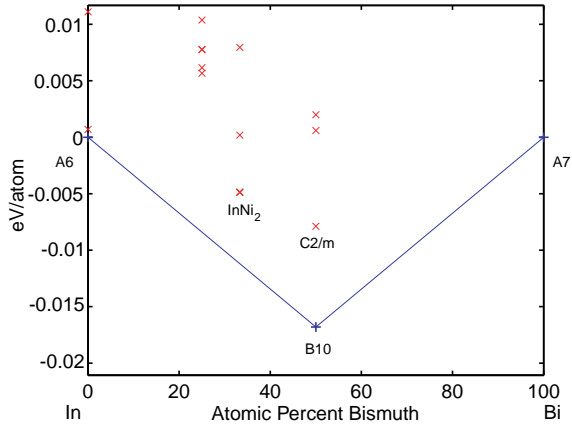,width=\picdim,clip=}
    \caption{Bi-In (Bismuth - Indium) ground state convex hull calculated in US-LDA.}
    \label{label.fig.BiInlda}
  \end{figure}
  \vspace{-1mm}
\end{center}
\vspace{-4mm}

{\it Due to the rather small energy differences and overall low
formation energies we further investigate the phase diagram with the
PAW-GGA potentials, as described in the method section. With PAW-GGA,
we find that BiIn-B10 has less negative formation energy, -6 meV/atom,
and therefore BiIn$_2$-InNi$_2$ becomes stable by having its energy
3.1 meV/atom below the tie-line In \twophase BiIn. The monoclinic
structure with space group C2/m \#12 at 50\% concentration is now
unstable having the formation energy of 10 meV/atom. The Bi$_3$In$_5$
phase still has positive formation energy (22 meV/atom) and remains
significantly higher the tie-line InNi$_2$ \twophase InBi (by 29
meV/atom).}  In summary, our results suggest that at low temperatures
BiIn$_2$ may be stable, while Bi$_3$In$_5$ is likely to be unstable.

\vspace{-2mm}
\begin{center}
  \vspace{-3mm}
  \begin{figure}
    \epsfig{file=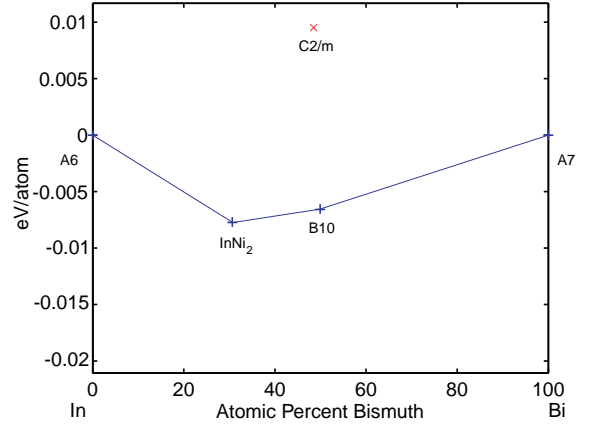,width=\picdim,clip=}
    \caption{Bi-In (Bismuth - Indium) ground state convex hull calculated in PAW-GGA.}
    \label{label.fig.BiIngga}
  \end{figure}
  \vspace{-1mm}
\end{center}

\begin{center}{\bf Bi-Mg (Bismuth - Magnesium)}\end{center}
Only one stable compound has been observed for the Mg-Sb system at low
temperatures: $\alpha$Bi$_3$Mg$_2$-La$_3$O$_2$
\cite{BB,PP,Wob62,Oh_92,Gru34}. Our calculations confirm the
low-temperature stability of this phase with a formation energy of
-252 meV/atom. We have not found any other stable compounds for this
system, therefore the experimental low temperature part of the diagram
is complete.

{\tablefont
  \begin{center}
    \begin{tabular}{||c||} \hline
      {\bf Bi-Mg system} \\ \hline
      Low Temperature Phases comparison chart \\ \hline
      \begin{tabular}{c|c|c}
        Composition  & Experimental  & \tablelineone                     \\
        \% Bi        & (\massalski)  & \tablelinetwo                     \\ \hline
        40           & D5$_2$        &  Bi$_3$Mg$_2$-D5$_2$ (\us)        \\
        \            &               & E$_f$=-252 meV/at.                 \\
      \end{tabular} \\ \hline
    \end{tabular}
  \end{center}
}

\vspace{-4mm}
\begin{center}
  \vspace{-3mm}
  \begin{figure}
    \epsfig{file=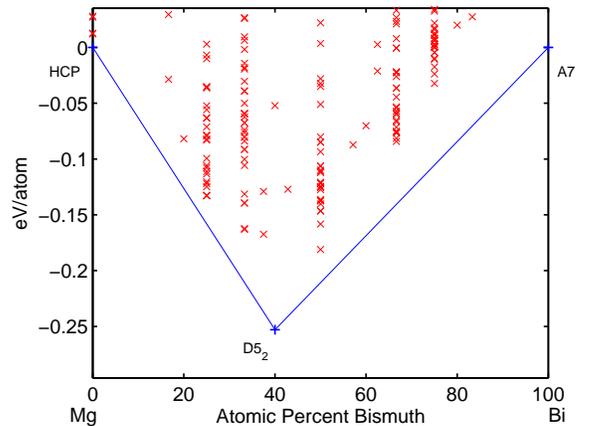,width=\picdim,clip=}
    \caption{Bi-Mg (Bismuth - Magnesium) ground state convex hull.}
    \label{label.fig.BiMg}
  \end{figure}
  \vspace{-1mm}
\end{center}

\begin{center}{\bf Bi-Sb (Bismuth - Antimony)}\end{center}
No intermediate stable compounds have been reported for the Bi-Sb
system and it is considered to be a non-compound forming system \cite{BB,PP}.
Our calculations confirm the absence of any stable phases. 
The structure with lowest formation energy in the whole range of
concentrations is found to be BiSb-B1 with E$_f$=21 meV/atom.


\begin{center}{\bf In-Mg (Indium - Magnesium)}\end{center}
Five ordered compounds have been reported for the system Mg-In at low
temperatures\cite{BB,PP,Hum38,Ray48,Hir68,Pic69,Fes76,Wat75}:
Mg$_{1.2}$In$_{2.8}$-L1$_2$ ($\gamma'$), MgIn-L1$_0$ ($\beta''$),
Mg$_2$In (prototype, $\beta_2$, Mg$_2$Tl in \cite{BB}),
Mg$_5$In$_2$-D8$_g$ ($\beta_3$), and Mg$_3$In ($\beta_1$) with Pearson
symbol hR48, and space group R$\bar{3}$m (\# 166) \cite{BB,PP} (note
that $\beta_1$ has Pearson symbol hR16 in \massalski). At high
temperatures phase $\beta_1$ transforms into the L1$_2$ structure,
which is similar to the high temperature field $\beta'$. Phases
$\beta_1$ and $\beta'$ share a eutectoid reaction at 202\DEG\,($\beta'
\leftrightarrow \beta_1+\beta_3$ at $\sim 27.5$\% of In), and a
perictectoid reaction at 337\DEG\,($\beta_1 \leftrightarrow \beta
+\beta'$ at $\sim 24$\% of In). Therefore we expect the morphology of
$\beta_1$ to be similar to L1$_2$ (with partial occupation) and to
find the structure Mg$_3$In-L1$_2$ under the tie-line In-A6 \twophase
MgIn-L1$_0$ at zero temperature.

{\tablefont
  \begin{center}
    \begin{tabular}{||c||} \hline
      {\bf In-Mg system} \\ \hline
      Low Temperature Phases comparison chart \\ \hline
      \begin{tabular}{c|c|c}
        Composition  & Experimental  & \tablelineone                     \\
        \% In        & (\massalski)  & \tablelinetwo                     \\ \hline
     $\sim$26 to 38.5& Mg$_3$In ($\beta_1$) & L1$_2$                     \\
        \            & hR48 R$\bar{3}$m \cite{PP}(low T), &             \\
        \            & $\sim$ L1$_2$(high T)    &                      \\ \hline  
        28.6         & Mg$_5$In$_2$-D8$_g$ ($\beta_3$)       & two phase region  \\
        \            &                            & D8$_g$ is 4.9 meV/at.\\
        \            &                            & above $\beta_1 \leftrightarrow \beta_2$ (\us) \\ 
        \            &                            & {\it D8$_g$ is 5.4 meV/at.}\\
        \            &                            & {\it above $\beta_1 \leftrightarrow \beta_2$ (\paw)} \\ \hline
        $\sim$34     & Mg$_2$In ($\beta_2$)   & Mg$_2$In stable        \\ 
        \            &                        & $\sim$3.5 meV/at. below      \\ 
        \            &                        & L1$_2$ \twophase L1$_0$ (\us) \\
        \            &                        & {\it $\sim$1 meV/at. below }    \\ 
        \            &                        & {\it L1$_2$ \twophase L1$_0$ (\paw) } \\ \hline
     $\sim$39 to 59  & MgIn-L1$_0$ ($\beta''$)   &  L1$_0$                     \\ \hline
        $\sim$69.5 to 75.5 &  Mg$_{1.2}$In$_{2.8}$-L1$_2$ ($\gamma'$) &  MgIn$_3$-L1$_2$ \\ 
      \end{tabular} \\ \hline
    \end{tabular}
  \end{center}
}

The off-stoichiometry Mg$_{1.2}$In$_{2.8}$-L1$_2$ phase
is not subject of our investigation, since it requires simulations
of disordered systems.
With our calculations, we find 
the ordered phase MgIn$_3$-L1$_2$ to be stable. Two
competing Long Period Superstructures of MgIn$_3$-L1$_2$, D0$_{23}$
and D0$_{24}$, are less favorable than the L1$_2$ phase by
$\sim$2.5 and $\sim$5.7 meV/atom, respectively.
This suggests that long-range interactions are weak in this
system at composition MgIn$_3$.
We confirm the
stability of MgIn-L1$_0$ ($\beta''$), Mg$_2$In-$\beta_2$, and
MgIn$_3$-L1$_2$ phases but find that Mg$_5$In$_2$-D8$_g$ lies 4.9
meV/atom above the tie-line Mg$_3$In-$\beta_1$ \twophase
Mg$_2$In-$\beta_2$. Besides, Mg$_2$In-$\beta_2$ is stable but with a
relative stability energy of 3.5 meV/atom below the tie-line
Mg$_3$In \twophase MgIn, which is within the error of
present calculations. 
We find a metastable compound, MgIn$_2$-C$_c$, to be $\sim$2.7 meV/atom above
the tie-line MgIn \twophase MgIn$_3$. This compound might  
become stable at higher temperatures and pressures.
{\it We recalculate the phase
diagram with PAW-GGA potentials and find essentially no difference
compared to the US-LDA results: Mg$_2$In-$\beta_2$ is 1 meV/atom below
the Mg$_3$In \twophase MgIn tie-line, Mg$_5$In$_2$-D8$_g$ is above the
Mg$_3$In-$\beta_1$ \twophase Mg$_2$In-$\beta_2$ by 5.4
meV/atom and MgIn$_2$-C$_c$ remains metastable by 4.6 meV/atom. 
In summary, our calculations confirm the
low temperature part of the Mg-In diagram known from experiment.
The results show that Mg$_5$In$_2$-D8$_g$ may not be stable at low
temperatures. In addition, we identify a metastable phase,
MgIn$_2$-C$_c$, which might be stable at higher pressures and temperatures. }

\begin{center}
  \begin{figure}
    \epsfig{file=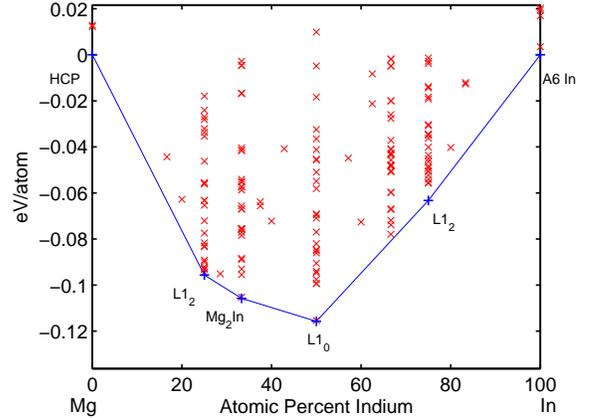,width=\picdim,clip=} 
    \caption{In-Mg (Indium - Magnesium) ground state convex hull.}
    \label{label.fig.InMg}
  \end{figure}
\end{center}

\newpage

\begin{center}{\bf In-Sb (Indium - Antimony)}\end{center}
The phase diagram of the system In-Sb is based on a
single compound
$\alpha$InSb-B3\cite{BB,PP,Pog49,Liu52,Hal63,Deg83,Gor83,Koz97}. It is
well studied at high pressures and temperatures where other
modifications of InSb ($\beta$InSb, $\gamma$InSb, and $\delta$InSb)
have been observed\cite{Deg82,Deg83}. The high-pressure transitions
have also been thoroughly studied with {\it ab initio}
methods\cite{Guo93,Kel00}. At zero pressure our calculations confirm
the experimental stability of the low temperature compound
$\alpha$InSb-B3. The InSb-B10 phase, stable for the similar binary
system In-Bi at 50\% concentration, here is unstable with an energy
$\sim$105 meV/atom higher than $\alpha$InSb-B3. We do not observe any
other stable compounds for the In-Sb system and conclude that the low
temperature experimental characterization of the system is complete.

{\tablefont
  \begin{center}
    \begin{tabular}{||c||} \hline
      {\bf In-Sb system} \\ \hline
      Low Temperature Phases comparison chart \\ \hline
      \begin{tabular}{c|c|c}
        Composition  & Experimental  & \tablelineone                     \\
        \% Sb        & (\massalski)  & \tablelinetwo                     \\ \hline
        50           &  B3           & $\alpha$InSb-B3                   \\
        \            &               & B10$\sim$105 meV/at.              \\
        \            &               & higher than B3 (\us)              \\ 
      \end{tabular} \\ \hline
    \end{tabular}
  \end{center}
}

\begin{center}
  \begin{figure}
    \epsfig{file=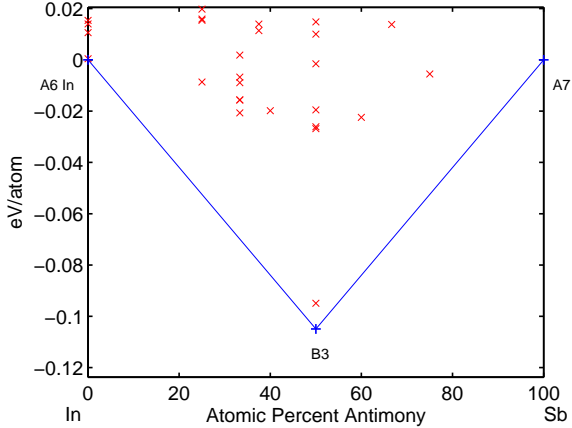,width=\picdim,clip=}
    \caption{In-Sb (Indium - Antimony) ground state convex hull.}
    \label{label.fig.InSb}
  \end{figure}
\end{center}


\begin{center}{\bf Mg-Sb (Magnesium - Antimony)}\end{center}

Similarly to the Bi-Mg system the Mg-Sb system has only one stable
compound at low temperature: $\alpha$Mg$_3$Sb$_2$-B5$_2$
\cite{PP,BB,Gru06,Gru34,Jon41,Bol62,Rao71}. Interestingly, there is no
solid solubility for Mg in Sb or Sb in Mg, although Mg and Sb have
similar dimensions (atomic radius ratio Sb/Mg=1.09) \cite{BB}.  At
high temperatures phase $\alpha$Mg$_3$Sb$_2$-B5$_2$ undergoes a
polymorphic change, as reported in \cite{Gru34}.  To our knowledge,
there is no information about the stability of
$\alpha$Mg$_3$Sb$_2$-B5$_2$ below 450\DEG\, and it is unclear how far
this stable phase extends into the low temperature region. Our
calculations confirm that this compound remains stable at low
temperatures having the formation energy of -404 meV/atom. No other
stable compounds have been found for this system, therefore the
experimentally known low-temperature part of the phase diagram is
complete.

{\tablefont
  \begin{center}
    \begin{tabular}{||c||} \hline
      {\bf Mg-In system} \\ \hline
      Low Temperature Phases comparison chart \\ \hline
      \begin{tabular}{c|c|c}
        Composition  & Experimental  & \tablelineone                      \\
        \% Sb        & (\massalski)  & \tablelinetwo                     \\ \hline
        40        & D5$_2$        &  Mg$_3$Sb$_2$-D5$_2$                         \\
      \end{tabular} \\ \hline
    \end{tabular}
  \end{center}
}

\vspace{-3mm}
\begin{center}
  \vspace{-3mm}
  \begin{figure}
    \epsfig{file=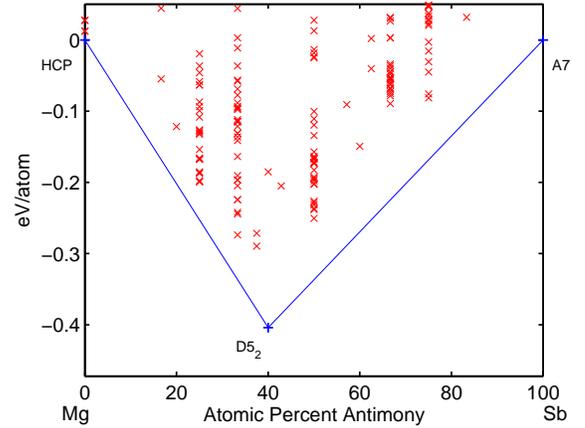,width=\picdim,clip=}
    \caption{Mg-Sb (Magnesium - Antimony) ground state convex hull.}
    \label{label.fig.MgSb}
  \end{figure}
  \vspace{-1mm}
\end{center}

\section{Conclusions}		
\label{section.conclusion}

The results of our systematic {\it ab initio} study of the six binary
systems Bi-In, Bi-Mg, Bi-Sb, In-Mg, In-Sb, and Mg-Sb suggest that the
experimentally known low-temperature phase diagrams for these alloys
are complete. Using the calculated ground state energies we complement
experimental information on the stability of several phases in the  
limit of low temperature and zero pressure: phase
$\alpha$Mg$_3$Sb$_2$-B5$_2$ is stable, phase Mg$_5$In$_2$-D8$_g$ may
not be stable and phase Bi$_3$In$_5$ is likely to be unstable.        
We find that compound MgIn$_2$-C$_c$, not observed experimentally,
is metastable but very close to stability and could therefore
exist at different temperatures and pressures.         
Further theoretical investigation of these
structures and search for other (meta)stable phases at finite
temperatures and pressures should include thermodynamic effects.

\section{Acknowledgments}
\label{section.acknowledgments}


This research has benefited from discussion with Gerbrand Ceder, Dane
Morgan, Milton Cole, Renee Diehl and Zi-Kui Liu.


\end{document}